\documentclass[11pt]{article}
\pdfoutput=1
\usepackage{amsmath}
\usepackage{amssymb}
\usepackage{graphicx,bbm,mathrsfs}
\usepackage{nicefrac}
\usepackage{slashed}
\usepackage{bbm}
\usepackage{geometry}
\usepackage{empheq}
\usepackage{stackrel}
\usepackage{ulem}
\usepackage{jheppub}
\usepackage{setspace}
\usepackage{relsize}
\usepackage{empheq}
\usepackage{pifont}
\usepackage{mathtools}
\usepackage{enumitem}
\usepackage{dcolumn}   
\usepackage{bm}        
\usepackage{graphicx,mathrsfs}
\usepackage{nicefrac}
\usepackage{multirow}
\usepackage{color}
\usepackage{mathtools}
\usepackage{makecell}
\usepackage{environ} 
\usepackage{lipsum} 
\usepackage{wasysym}
\usepackage{hhline,colortbl}  
\usepackage{tikz}
\usepackage{graphicx}
\usepackage{tensor}
\usepackage{dsfont}
\usepackage{simpler-wick}

\usepackage{booktabs}
\usepackage{mdframed}
\usepackage{xcolor}
\definecolor{EqFrame}{RGB}{235,245 ,250 }

 \NewEnviron{Smaller11}{
           \scalebox{1.1}{$\BODY$} 
 } 
 \NewEnviron{Smaller08}{
           \scalebox{0.8}{$\BODY$} 
 }

\newcommand{\be}{\begin{equation}}
\newcommand{\ee}{\end{equation}}
\newcommand{\bea}{\begin{eqnarray}}
\newcommand{\eea}{\end{eqnarray}}

\newcommand{\Zvec}{\bm Z}

\newcommand{\M}{M}
\newcommand{\Q}{Q}
\newcommand{\e}{e}

\renewcommand\labelenumi{(\roman{enumi})}
\renewcommand\theenumi\labelenumi

\usepackage{titlesec}

\titleformat*{\section}{\Large\bfseries}
\titleformat*{\subsection}{\large\bfseries}
\titleformat*{\subsubsection}{\large\bfseries}
\titleformat*{\paragraph}{\large\bfseries}
\titleformat*{\subparagraph}{\large\bfseries}

\makeatletter
\newcommand*{\prodsym}{%
  \DOTSB
  \mathop{
    \mathchoice
      {\rlap{\kern.3em\rotatebox[origin=c]{-90}{}}{\prod}}
      {\vcenter{\rlap{\kern.2em\rotatebox[origin=c]{-90}{}}}{\prod}}
      {\sum}{\sum}
  }\slimits@
}
\makeatother

\makeatletter
\DeclareFontFamily{OMX}{MnSymbolE}{}
\DeclareSymbolFont{MnLargeSymbols}{OMX}{MnSymbolE}{m}{n}
\SetSymbolFont{MnLargeSymbols}{bold}{OMX}{MnSymbolE}{b}{n}
\DeclareFontShape{OMX}{MnSymbolE}{m}{n}{
    <-6>  MnSymbolE5
   <6-7>  MnSymbolE6
   <7-8>  MnSymbolE7
   <8-9>  MnSymbolE8
   <9-10> MnSymbolE9
  <10-12> MnSymbolE10
  <12->   MnSymbolE12
}{}
\DeclareFontShape{OMX}{MnSymbolE}{b}{n}{
    <-6>  MnSymbolE-Bold5
   <6-7>  MnSymbolE-Bold6
   <7-8>  MnSymbolE-Bold7
   <8-9>  MnSymbolE-Bold8
   <9-10> MnSymbolE-Bold9
  <10-12> MnSymbolE-Bold10
  <12->   MnSymbolE-Bold12
}{}

\let\llangle\@undefined
\let\rrangle\@undefined
\DeclareMathDelimiter{\llangle}{\mathopen}%
                     {MnLargeSymbols}{'164}{MnLargeSymbols}{'164}
\DeclareMathDelimiter{\rrangle}{\mathclose}%
                     {MnLargeSymbols}{'171}{MnLargeSymbols}{'171}
\makeatother

\begin{document}

\vspace*{4mm}

\thispagestyle{empty}

\begin{center}

\begin{minipage}{20cm}
\begin{center}
\hspace{-5cm }
\LARGE
\sc
On The Black Hole Weak Gravity Conjecture 
 \\  
\hspace{-5cm }    
and Extremality in the Strong-Field Regime
\end{center}
\end{minipage}
\\[30mm]

\renewcommand{\thefootnote}{\fnsymbol{footnote}}

\renewcommand{\thefootnote}{\fnsymbol{footnote}}

{\large  
Sergio~Barbosa$^{\, a}$ \footnote{sergio.barbosa@aluno.ufabc.edu.br}\,, 
Sylvain~Fichet$^{\, a}$ \footnote{sylvain.fichet@gmail.com}\,, 
Lucas~de~Souza$^{\, b}$ \footnote{souza.l@ufabc.edu.br} 
}\\[12mm]
\end{center} 
\noindent

\indent \; ${}^a\!$ 
\textit{CCNH, Universidade Federal do ABC,} \textit{Santo Andr\'e, 09210-580 SP, Brazil}

\indent \; ${}^b\!$ 
\textit{CMCC, Universidade Federal do ABC,} \textit{Santo Andr\'e, 09210-580 SP, Brazil}
\\

\addtocounter{footnote}{-3}

\vspace*{8mm}
 
\begin{center}
{  \bf  Abstract }
\end{center}
\begin{minipage}{15cm}
\setstretch{0.95}

We point out that the Weak Gravity Conjecture (WGC) implies that sufficiently small extremal black holes are necessarily in the strong-field regime of electrodynamics, and therefore probe the UV completion of the Maxwell sector. 
To investigate the WGC bounds arising from these small extremal black holes, we revisit black hole decay in generic field theories in asymptotically flat space. 
We derive a necessary  and a sufficient condition for any black hole to decay, the latter amounting to a bound on the growth of charge relative to mass.
We apply these  conditions to extremal black holes derived in various UV completions of the Maxwell sector. We find that the Euler-Heisenberg and DBI effective actions satisfy the sufficient condition for decay, while the ModMax model fails the necessary one, rendering it incompatible with the WGC.
Using  the decay conditions, we show that the black hole WGC implies positivity of the $U(1)$ gauge coupling beta function. This provides an independent argument that classically stable (embedded-Abelian) colored black holes cannot exist. We also show that the black hole WGC constrains conformal hidden sector models, and is always satisfied in their AdS dual realizations.

    \vspace{0.5cm}
\end{minipage}

\newpage
\setcounter{tocdepth}{2}

\tableofcontents

\newpage 
\section{Introduction \label{se:Intro}}

Extremal and near-extremal black holes can be viewed as probes of high-energy physics.  One reason for this is  the strong electromagnetic fields they emit near the horizon (see e.g.~\cite{Maldacena:2020skw}).\,\footnote{
Ultraviolet (UV) sensitivity also manifests at the level of the black hole tidal perturbations, see
\cite{Horowitz:2023xyl, Horowitz:2024dch,Barbosa:2025uau}. }  
Another reason is that, from the perspective of the UV completion of quantum gravity, black holes may be considered as semiclassical descriptions of super-Planckian elementary states \cite{Susskind:1993ws,Horowitz:1996nw}. In this context, extremal black holes provide an arena for conjectures about quantum gravity, see e.g. \cite{ArkaniHamed:2006dz,Palti:2019pca,Harlow:2022ich}. 
This note examines the interplay between extremal black holes, strong fields and  the Weak Gravity Conjecture (WGC). 

Beyond the statement that gravity is the weakest force, the WGC encompasses a set of 
assertions about the consistency of  field theories with quantum gravity \cite{Palti:2019pca,Harlow:2022ich}. 
Many versions of the WGC are motivated from the ultraviolet (UV), drawing on  a large number of examples from string theory.  
Arguments  from the infrared (IR) also exist, however, see \cite{Kats:2006xp,
Cheung:2014ega,
Endlich:2017tqa,
Cheung:2018cwt,
Hamada:2018dde,
Loges:2019jzs,Goon:2019faz,Jones:2019nev, Chen:2019qvr, Bellazzini:2019xts,Loges:2020trf,Arkani-Hamed:2021ajd,Cao:2022iqh,DeLuca:2022tkm, Bittar:2024xuc, Knorr:2024yiu}. 
One of the original IR arguments for the WGC 
is the idea that extremal black holes of any size must be able to decay \cite{ArkaniHamed:2006dz}. 

The central hypothesis of this note is that all extremal black holes must be able to decay in any field theory consistent with quantum gravity. 
Even though there are compelling arguments for it, it remains a conjecture, which we refer to as the \textit{black hole WGC}.

How can the black hole WGC hold in   a theory consisting only of  photons and gravitons? With no particles available to dissipate charge, extremal black holes can only decay into smaller black holes. Surprisingly, such decay would be impossible in pure General Relativity (GR) with Maxwell electromagnetism. 
One statement of this property is  that extremal black holes of any size have charge-to-mass ratio  equal to one. 

The solution to the apparent tension between  the black hole WGC and the pure  GR-Maxwell theory is that a field theory emerging from quantum gravity is a low-energy  limit, implying that it generally deviates  from both GR and Maxwell electromagnetism.\,\footnote{
The  low-energy theory may arise as limit of either the theory of quantum gravity itself or of some intermediate sub-Planckian theory. Strictly speaking,  only the former can be referred to as true UV completion while the latter is rather an ``intermediate'' UV completion. For convenience we   refer to any theory arising  immediately  above a UV cutoff as the UV completion. 
}
 These deviations 
 can be such that extremal black holes are allowed to decay.

The condition for extremal black hole decay is conveniently expressed in terms of the black hole charge-to-mass ratio $Z$, defined as 
 \be
 Z=\frac{\sqrt{2}}{\kappa}\frac{Q_{\displaystyle \circ}}{M_{\displaystyle \circ}}
 \ee
 where $Q_{\displaystyle \circ}$, $M_{\displaystyle \circ}$ are the physical charge and mass of the black hole, computed by integrals at spatial infinity in asymptotically flat spacetime. 
We denote the charge and a charge-to-mass ratio of an extremal black hole as 
$\bar Q\equiv Q_{\rm extremal}$, $ \bar Z\equiv Z_{\rm extremal}$
throughout this note.

\subsubsection*{Black hole WGC beyond EFT}

In the effective field theory (EFT) regime, the deviations from GR and Maxwell electromagnetism are encoded in a few irrelevant operators, such that the deviations to the charge-to-mass ratio take the schematic form  $\bar Z=1+\frac{a}{M^n}$, with $n=2$ in $d=4$ spacetime dimensions. 
The condition for extremal black hole  decay is usually expressed as the condition that $a>0$ \cite{Kats:2006xp}, i.e.\,\footnote{In the nomenclature of \cite{Harlow:2022ich} this is the {\it mild} form of the WGC, here applied to deviations from $d=4$ GR.   }
\be \bar Z>1 \label{eq:Zusual} \,.\ee 
In the EFT context, \eqref{eq:Zusual} is equivalent to $\bar Z$ being a decreasing function of $M$.

In this note, we derive conditions 
for the decay of any extremal black holes (i.e. the black hole WGC)  that hold {beyond} the EFT regime,  for any gravitational field theory in asymptotically flat spacetime.  
\global\mdfdefinestyle{EqFrame}{ linecolor=white,linewidth=3pt,
backgroundcolor=EqFrame,
leftmargin=0cm,rightmargin=0cm }
\begin{mdframed}[style=EqFrame]
 \textbf{Conditions for extremal black hole decay  }\\~
\\
{Sufficient condition:}
\be
\shortstack[l]{ \it The extremality curve satisfies  $\displaystyle \frac{\partial\, \bar Z}{\partial M}<0$
\textrm{\it for all} $M$. } \label{eq:WGC_sufficient_condition}
\ee
{Necessary condition:}
\be {\shortstack[l]{
\textrm{
\it There is a discrete subset of masses {$A$}
 such that   $\bar Z\big|_A$  is strictly decreasing. }
}} \label{eq:WGC_necessary_condition}
\ee
\end{mdframed}
These conditions are shown  in App.\,\eqref{se:Growth_bound}, for both single and multiple $U(1)$. 

The sufficient condition \eqref{eq:WGC_sufficient_condition}, viewed in terms of black hole charge, limits  the charge growth as a function of mass, hence we refer to it as the \textit{charge growth bound}.\,\footnote{A recent work \cite{Abe:2025vdj} has derived a similar monotonicity property from causality constraints in the electromagnetic sector, in the regime where gravitational corrections are neglected.} 
The trick to derive \eqref{eq:WGC_sufficient_condition} is to observe that the extremal black hole charge $\bar Q$ is a {subadditive} function of $M$.
The necessary condition \eqref{eq:WGC_necessary_condition}
 holds even if $\bar Z$ is {non-monotonic} in $M$. In fact, this condition holds even if black holes masses are quantized, which is a QFT-motivated possibility considered in e.g. \cite{Dvali:2011nh}.

\subsubsection*{Black hole WGC and strong fields}

Since conditions \eqref{eq:WGC_sufficient_condition}, \eqref{eq:WGC_necessary_condition} apply beyond the EFT regime, they provide results that would otherwise be inaccessible.   The specific beyond-EFT regime of our focus  is the \textit{strong-field regime} of electrodynamics. 
Using  the necessary condition \eqref{eq:WGC_necessary_condition}, we  will show  that  certain extensions of the Maxwell sector that are probed at strong field are  incompatible with the black hole WGC --- and thus belong  to the ``swampland'' of low-energy gravitational field theories.

In this note, the interplay between strong-field electrodynamics and extremal black holes is two-fold.  We will show that sufficiently small extremal black holes generically experience a strong-field regime near their horizon. To obtain this claim, we  apply the WGC to \textit{large} extremal black holes. This is an application distinct from the WGC application  to small black holes described in the previous paragraph. 
Hence our analysis involves two separate connections between the black hole WGC and the strong-field regime: applying the WGC to large black holes implies strong fields near small black holes, while applying it to small black holes constrains the Maxwell UV completions.

\subsubsection*{Outline}

Our focus  is on non-spinning charged black holes in asymptotically flat space. Our analysis proceeds in the following steps.
 In section  \ref{se:EBH_UVProbes}, we  discuss the various EFT scales that appear in the extremal black hole background. We point out that two field strength regimes exist depending on the black hole size, and that this phenomenon is ensured by the WGC.  
We then apply our {conditions for extremal black hole decay} 
to various nonlinear QED models in section \ref{se:QED}: Euler-Heisenberg, Dirac-Born-Infeld, and the so-called ModMax model. Finally, in section \ref{se:BetaApplications}, we  relate the black hole WGC  to positivity of the $U(1)$ beta function  and study the consequences of this fact for colored black holes and conformal hidden sector models.  Section \ref{se:conclusion} summarizes. 
Appendix \ref{se:Growth_bound} contains the proofs of conditions \eqref{eq:WGC_sufficient_condition}, \eqref{eq:WGC_necessary_condition} for single and multiple charges. Details on magnetic black holes are given in appendix  \ref{app:BH}.

\section{Extremal Black Holes as UV Probes}

\label{se:EBH_UVProbes}

In this section we use dimensional analysis to show that sufficiently small  extremal black holes can exhibit a strong-field regime. We then point out that the existence of this regime is implied by the black hole WGC itself when applied in the weak-field regime.

\subsection{Effective Action and EFT Scales}

The dynamics of spacetime and matter at distances larger than the Planck length can be described by an Effective Field Theory (EFT). 
In the presence of a $U(1)$ gauge field, the quantum effective action $\Gamma$ encoding all the information  is built from the Riemann tensor, the $U(1)$ field strength and their covariant derivatives, $\Gamma=\Gamma[R_{\mu\nu\rho\sigma},F_{\mu\nu},\nabla]$. 
The effective action can be organized with respect to the physical scales associated to each of the  building blocks:
\be
\Gamma[R_{\mu\nu\rho\sigma},F_{\mu\nu},\nabla] = 
\int d^4 x \sqrt{-g} \left[ \frac{1}{2\kappa^2}R-\frac{1}{4 \e^2 } F_{\mu\nu}F^{\mu\nu}+{\cal O}\left( \frac{{\rm Riem}^2}{\Lambda_R^4} ,\frac{F^4}{\Lambda^8_F} , \frac{\nabla}{\Lambda} \right)  \right]\,.
\label{eq:Gamma_exp}
\ee
Here $\Lambda_R$ controls the spacetime curvature expansion,  $\Lambda_F$ controls the field strength expansion, and $\Lambda$ is a mass scale controlling the derivative expansion.  
In the  gravitational sector, derivatives can be converted into curvature. Hence in that sector a single scale controls both curvature and derivative expansions. The derivatives in \eqref{eq:Gamma_exp} are those acting on matter fields. 

The pure Maxwell sector of the action is encoded in a Lagrangian  denoted ${\cal L}_F$, with  $\int d^4 x\sqrt{-g} {\cal L}_F\subset \Gamma$. We assume that ${\cal L}_F={\cal L}_F[F^2,F\tilde F]$ is analytical in  $F^2$, $F\tilde F$ and vanishes for $F^2\to 0$, $F\tilde F\to0$. Whether or not  ${\cal L}_F$ can be truncated to its first effective operators is a key distinction explored throughout this note.

\subsection{EFT Scales Near the Horizon }
 
Consider charged black holes solutions  in asymptotically flat space with physical mass $ M_{\displaystyle \circ}\equiv 4\pi \M$ and fractional charge $Q_{\displaystyle \circ}\equiv 4\pi \Q>0$, computed from integrals at spatial infinity (see App.\,\ref{app:BH}). We consider that the black hole is an approximated  Reissner-Nordstr\"om solution (RN) such that it has two horizons with radii given by
\begin{equation}
    r_\pm \approx \frac{1}{2}\left(
 \kappa^2 \M \pm\kappa\sqrt{\kappa^2\M^2-2\Q^2}
 \right)\,.
\end{equation}
 This approximation holds in the EFT regime of the effective action. 

The charged black holes are affected by the three type of corrections listed in \eqref{eq:Gamma_exp}. Near the outer horizon $r_+$, the building blocks appearing in \eqref{eq:Gamma_exp} are  estimated as\,\footnote{Throughout this section we ignore  the numerical factors in the estimates.  }
\be
\nabla \sim \frac{1}{r_+}\,,\quad \quad {\rm Riem} \sim \frac{1}{r_+^2}, \,\quad\quad F_{\mu\nu}\sim \frac{\e \Q}{r_+^2}\,. 
\ee
The corresponding expansion parameters
showing up in \eqref{eq:Gamma_exp} can be written as
 \be
\frac{\square}{\Lambda^2 } \sim \frac{1}{\Lambda^2 r^2_+}\equiv\epsilon , \,\quad\quad
\frac{\rm Riem}{\Lambda_R^2} \sim \frac{1}{\Lambda_R^2 r_+^2} \equiv\epsilon_R\,,\quad  \quad 
\frac{F^2}{\Lambda_F^4}\sim \frac{\e^2\Q^2 }{ \Lambda_F^4 r^4_+ } \equiv\epsilon_F\,.
\label{eq:expansion_parameters}
 \ee
Here  we choose same number of derivatives in each numerator so that the $\epsilon$ parameters can be directly compared to each other. 
 The    $\epsilon$ and $\epsilon_R$ parameters are analogous.
 The $\epsilon_F$ parameter crucially differs from these due to the dependence on the black hole charge.  

\subsubsection{The weak and strong-field regimes}

\label{se:two_regimes}
 
The expansion series can be truncated if  $\epsilon_i<1$,  schematically, which translates as the conditions
 \be\Lambda,\Lambda_R > \frac{1}{r_+}\,,\quad  \quad  
 \Lambda_F > \frac{\sqrt{\e \Q}}{r_+}\equiv \Lambda_F^c\,. \label{eq:expansion_conditions}
 \ee
   Consider $\Lambda_F<\Lambda,\Lambda_R$, which is the most important case as we will see further below. 
{For radii $r_+   > \frac{1}{\Lambda },\frac{1}{\Lambda_R}$ the higher derivative and higher curvature terms can be neglected. In contrast, 
    the field strength expansion condition  $ \Lambda_F > \Lambda_F^c$ is \textit{not} necessarily satisfied. Namely, at fixed $r_+$, the $ \Lambda_F > \Lambda_F^c$ condition  does not hold for a sufficiently charged black hole.

 This demonstrates that there exist two  regimes
  for the electromagnetic field  of the charged black hole, that are separated by the critical value $\Lambda_F^c$. Expressed in terms of $\Lambda_F$, the $ \Lambda_F > \Lambda_F^c$ case corresponds to the \textit{weak-field} regime, in which the most important operators are the $F^4$ ones. The $ \Lambda_F < \Lambda_F^c$ case corresponds to the \textit{strong-field} regime for which the entire power series of field strength  must be taken into account. 
This latter case depends on the UV completion of the Maxwell sector encoded in  ${\cal L}_F$.

\subsubsection{Extremal limit}

\label{se:EBH_scales}

The emergence of a strong-field regime is most pronounced for extremal black holes.
We denote the  extremal radius as $r_+=r_-\equiv r_h$, with the mass and charge  related by $\M \approx  \frac{2r_h}{\kappa^2}$, $\Q\approx \frac{\sqrt{2}r_h}{\kappa}$. For an extremal black hole, the field strength at the horizon reaches $F_{\mu\nu}\sim \frac{\e}{\kappa r_h}$. The critical scale of the strong-field regime is then $
\Lambda^c_F \sim \sqrt{\frac{\e}{r_h \kappa}} 
$. 
This implies that, for a given field strength expansion scale $\Lambda_F$, extremal black holes are in the strong-field regime when   their  radius is \textit{smaller} than the critical value $r_h^c $ given by 
\be r_h^c\equiv \frac{\e}{\kappa \Lambda^2_F} \,. \ee

\subsection{Strong-Field Regime from the WGC }

In the extremal limit, the curvature and field strength expansion parameters take the form 
\be
\epsilon_R=\frac{1}{\Lambda_R^2 r_h^2}\,,\quad \quad \epsilon_F= \frac{\e^2}{\Lambda_F^4 \kappa^2 r_h^2}\,. 
\ee
The competition between these two parameters turns out to be decided by the weak gravity conjecture.

\subsubsection{The WGC positivity bound on the EFT}

Let us consider extremal black holes in the weak-field regime, i.e. $r_h>r_h^c$. In that regime the effective action can be truncated to the first leading terms, 
\be \Gamma=\int d^4x \sqrt{-g} \left({\cal L}_{\rm EFT} + {\cal O}(F^6,RF^4, R^2 F^2, R^3)\right)\ee
where the leading operators can be reduced to 
\begin{align}
{\cal L}_{\rm EFT} =     \frac{1}{2\kappa^2}{R} - \frac{1}{4\e^2 }F_{\mu \nu}F^{\mu \nu}  + \gamma_1 R^{\mu \nu \rho \sigma}F_{\mu \nu}F_{\rho \sigma} + \gamma_2 (F_{\mu \nu}F^{\mu \nu})^2 +  \gamma_3  (F_{\mu \nu}\tilde F^{\mu \nu})^2    
\label{eq:EinsteinMawxellEFT}
\end{align}
 using standard EFT techniques \cite{Manohar:2018aog,Bittar:2024xuc}.

These  operators induce a slight deviation to the extremality curve, that has been computed in a number of references, see e.g. \cite{Kats:2006xp,  Jones:2019nev, Barbosa:2025uau} and also \cite{DeLuca:2022tkm, Barbosa:2025uau} for higher order. 
The result for pure electric charge and pure magnetic charges are respectively  
\be \bar Z_{e,m}=1\pm \frac{2\e^2}{5 r_h^2}\gamma_1+\frac{8\e^4}{5 \kappa^2 r_h^2}\gamma_2\,. 
\ee
Applying either  the EFT-level black hole WGC condition  \eqref{eq:Zusual},  or the more general condition \eqref{eq:WGC_sufficient_condition}  
(using that $r_h\frac{d }{d r_h}\approx M\frac{d }{d M}$)
to both electric and magnetic cases leads to a positivity bound on the deviation to extremality, 
\,\footnote{The bound \eqref{eq:WGC_positivity} is reproduced using unitarity of forward amplitudes in a regularized approach to the graviton $t$-channel singularity \cite{Bellazzini:2019xts}. This illustrates the connection between IR consistency bounds and the black hole WGC.
That \eqref{eq:WGC_positivity} is exactly obtained from \cite{Bellazzini:2019xts} is perhaps surprising since unitarity bounds tend to weaken in gravitational EFTs, see e.g. \cite{Alberte:2020jsk,Alberte:2020bdz, Caron-Huot:2021enk, Hamada:2023cyt,Chang:2025cxc}. }
\be
4\e^2\gamma_2  - \kappa^2 |\gamma_1| >0 \,.
\label{eq:WGC_positivity}
\ee

\subsubsection{Existence of the strong-field regime}

We combine the WGC bound at weak field \eqref{eq:WGC_positivity}
 with the dimensional analysis made throughout section \ref{se:EBH_UVProbes} by identifying $\gamma_1\sim \frac{1}{\Lambda_R^2}$, $\gamma_2\sim \frac{1}{\Lambda_F^4}$.\,\footnote{This identification assumes no unnaturally small coefficients in the Wilsonian sense. See also \cite{Cheung:2014vva} for naturalness considerations in the WGC context.  } 
  Dropping the numerical factor,  the bound  becomes 
\be
\frac{\Lambda_R}{\kappa} > \frac{\Lambda_F^2}{\e}\,.\label{eq:WGC_EFT}
\ee
This bound can be interpreted as a version of  the statement that gravity is the weakest force, made at the level of the EFT scales introduced in \eqref{eq:Gamma_exp}.

At the level of the expansion parameters, \eqref{eq:WGC_EFT}
 translates as the hierarchy
  \be \epsilon_F>\epsilon_{R}
  \label{eq:epsilon_hierarchy}
  \,.\ee 
Importantly, this condition holds for any black hole radii.

For $\epsilon_R <1$, i.e. in the validity domain of the gravitational EFT, the inequality \eqref{eq:epsilon_hierarchy}  implies the existence of  both a  weak-field regime, where  $\epsilon_R, \epsilon_F <1$, and a strong-field regime, where $\epsilon_R < 1$ but $\epsilon_F >1$. 

We thus conclude that the WGC applied to extremal black holes in the weak-field regime automatically implies the existence of extremal black holes in the   strong-field regime. This strong-field regime is absent only if the WGC inequality \eqref{eq:WGC_EFT} is saturated --- or if Wilsonian naturalness breaks down  such that the identification $\gamma_1\sim \frac{1}{\Lambda_R^2}$, $\gamma_2\sim \frac{1}{\Lambda_F^4}$ is not valid anymore.

\subsubsection{On Maxwell dominance }

\label{se:Maxwell_dom}

The inequality $\epsilon_F>\epsilon_{R}$ caused by the  WGC has practical consequences. If there is a hierarchy $\epsilon_F\gg \epsilon_{R}$, then the gravitational corrections to the extremal black hole are small with respect to the Maxwell corrections. This is very useful because, in practice, the Maxwell effective action is easier to compute than the gravitational sector, see e.g. the full Euler-Heisenberg Lagrangian. Moreover the computations at the level of the black hole metric are also simplified, as we will see in next section. 
The beauty of small extremal black holes is that the electromagnetic field is so strong that gravity corrections can be neglected. 

This phenomenon can be seen, for example, at the level of the extremality relation, or in  the Love numbers of extremal black holes, as discussed in  \cite{Barbosa:2025uau}.

 \begin{figure}[t]
     \centering
    \includegraphics[width=0.99\linewidth,trim={0.5cm 6.5cm 1cm 7cm},clip]{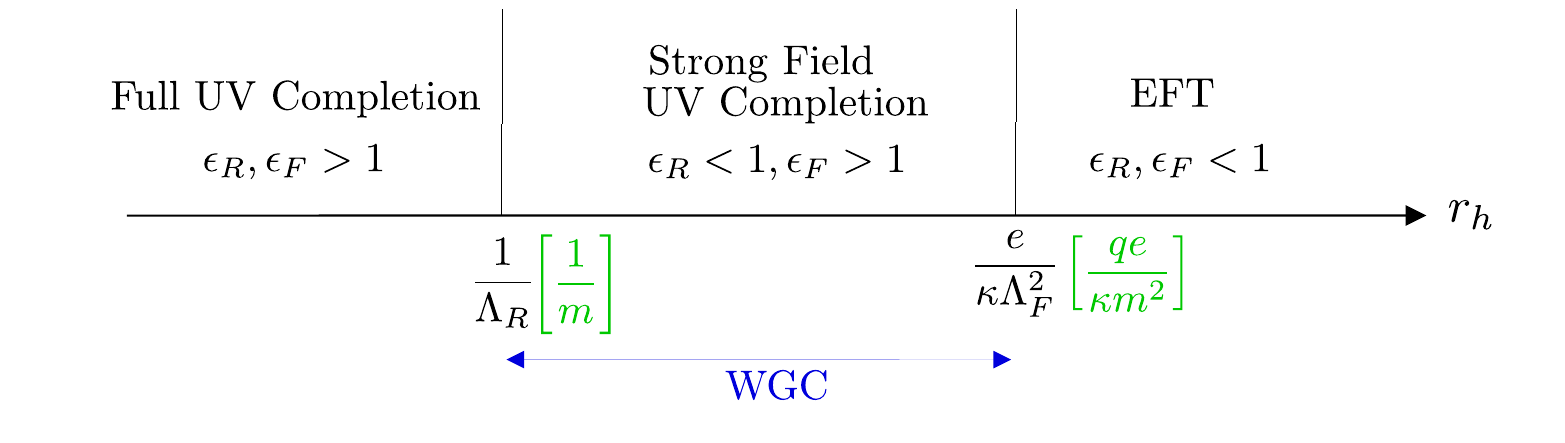}
\caption{     \label{fig:Sketch_Large_Field} 
The  regimes of the effective action in the extremal black hole background. We assume $\Lambda=\Lambda_R$. The values in brackets correspond to the Euler-Heisenberg case. The WGC ensures the existence of the intermediate strong-field domain.  } 
 \end{figure}

\subsubsection{Summary}

We summarize the points obtained through this section.

Near the extremal black hole horizon, the three expansion parameters $\epsilon, \epsilon_R, \epsilon_F$ have same dependence in $r_h$,  hence their competition is controlled by the hierarchy between the EFT scales, independently of the black hole radius. 
The  derivative expansion  scale $\Lambda$ is in general unrelated to the two others. For our purposes, it is enough to assume that it is identical to the curvature expansion scale, $\Lambda\sim \Lambda_R$, such that $\epsilon=\epsilon_R$.

We have found that the WGC, when applied to extremal black holes in the weak-field regime,  implies $\epsilon_F>\epsilon_{R}$. 
This hierarchy  implies that the quantum effective action $\Gamma$ experiences   
 {three} regimes. 
\begin{itemize}
    \item The $\epsilon_R,\epsilon_F<1$ is  the EFT regime for which $F^4$, $RF^2$, $R^2$ operators are the most important ones.  
 
    \item The $\epsilon_R<1, \epsilon_F>1$ case we refer to as the \textit{strong-field UV completion}. In this regime the derivative expansion of $\Gamma$ can still be truncated while the field strength expansion cannot.  The UV completion of the Maxwell sector encoded in ${\cal L}_F$ must be taken into account. 

        \item The $\epsilon_R,\epsilon_F>1$ case is the full UV completion, that would need a UV completion of both gravitational and Maxwell sectors. 
\end{itemize}

 These conclusions  are summarized in Fig.\,\ref{fig:Sketch_Large_Field}.  
The strong-field UV completion contains some information about the full UV completion, but not all of it. In principle, it is possible for two different full UV completions to give rise to the same strong-field UV completion once the derivative expansion is truncated.

\subsection{Example: Charged Particles}

A central example is the one of the massive charged particle. Integrating out exactly a  massive charged particle of mass $m$ and fractional charge $q>0$ leads  
to an example of effective action $\Gamma[R_{\mu\nu\rho\sigma},F_{\mu\nu},\nabla]$ considered through this section. 

The first terms of the derivative expansion have been thoroughly computed via the heat kernel method, see e.g. \cite{Vassilevich:2003xt}. At leading order of the derivative expansion, the field strength expansion corresponds to the Euler-Heisenberg i.e.~nonlinear QED Lagrangian \cite{Dunne:2004nc}, that we discuss in more details in \ref{se:EH}.
 The direct computation makes clear that the three EFT scales controlling the derivative, curvature, and field strength expansions defined in \eqref{eq:Gamma_exp} are
\be
\Lambda\sim m\,\quad \quad \Lambda_R\sim m\,,\quad \quad \Lambda_F \sim \frac{m}{\sqrt{ q}}\,. \label{eq:Lambda_charged_particle}
\ee

Substituting \eqref{eq:Lambda_charged_particle} into \eqref{eq:expansion_parameters} we have $\epsilon=\epsilon_R=\frac{1}{(m r_h)^2}$. This means that the derivative and curvature expansions can be truncated if the particle Compton wavelength is much smaller than the black hole radius. 
On the other hand, the field strength expansion parameter is
\be
\epsilon_F=\frac{\e^2 q^2}{\kappa^2 m^4 r_h^2} \,.  \label{eq:epsilonF_particle}
\ee
This implies that  the critical radius is $r^c_h=\frac{\e q}{\kappa m^2}$. 
An extremal black hole with radius smaller than $r_h<r^c_h$ is in the strong-field regime. 
Similar observations have been done in  \cite{Abe:2023anf}. 

The scales identified in this section are summarized in Fig.\,\ref{fig:Sketch_Large_Field}. 
As a sanity check of our  analysis, one can  verify explicitly in section \ref{se:QED} that $\epsilon_F$ corresponds precisely to the expansion parameter of the Euler-Heisenberg Lagrangian. 

Finally, one may notice that applying the WGC inequality \eqref{eq:WGC_EFT} to the EFT scales \eqref{eq:Lambda_charged_particle} produces the bound $\frac{\e q}{\kappa m}\gtrsim 1$. This is the particle version of the WGC \cite{Palti:2019pca}.\,\footnote{The connection between the particle WGC and  both black hole WGC and infrared consistency has been studied \cite{Cheung:2014ega, 
Hamada:2018dde, Bellazzini:2019xts,
Chen:2019qvr, 
Arkani-Hamed:2021ajd, 
Bittar:2024xuc,
Knorr:2024yiu}.   This connection does not hold in certain spacetime dimension $d\neq 4$, while the connection between  black hole WGC and infrared consistency remains intact under changes of dimension \cite{Bittar:2024xuc}. 
}

\section{Black Hole WGC and  Nonlinear QED Models} \label{se:QED}

\subsection{Extremal Black Hole Beyond the Weak-Field Regime}\label{se:gen}

In section \ref{se:EBH_UVProbes} we have identified qualitatively the strong-field regime of extremal black holes. In this section we compute explicit results in this regime.

Starting from the general quantum effective action $\Gamma[R_{\mu\nu\rho\sigma},F_{\mu\nu},\nabla]$, we expand in curvature and derivatives, but keep the full Lagrangian for the field strength, denoted ${\cal L}_F$:  
\be
\Gamma[R_{\mu\nu\rho\sigma},F_{\mu\nu},\nabla] = 
\int d^4 x \sqrt{-g} \left[ \frac{1}{2\kappa^2}R +{\cal L}_F[F^2, F\tilde F]+{\cal O}\left( \frac{{\rm Riem}^2}{\Lambda_R^4} , \frac{\nabla}{\Lambda} \right)  \right]\,. 
\label{eq:Gamma_F}
\ee
where $F^2=F_{\mu\nu}F^{\mu\nu}$, $F\tilde F=\frac{1}{2}\epsilon^{\mu\nu \rho\sigma}F_{\mu\nu}F_{\rho\sigma}$. 

We consider spherically symmetric black hole solutions
\begin{equation}
    ds^2 = - f_t(r) dt^2 + \frac{1}{f_r(r)} dr^2 + r^2 d\Omega^2\,.
    \label{eq:metric_gen}
\end{equation}
Details of the solving are given in App.\,\ref{app:BH}.  
We find $f_t(r) = f_r(r) \equiv f(r)$ in the absence of corrections to gravity, with 
\be
f(r) = 1 - \frac{\kappa^2 M}{r} - \frac{\kappa^2}{r}\int_r^\infty dr^\prime r^{\prime 2} \mathcal{L}_{F}(r^\prime)\,. \label{eq:f_gen}
\ee

For our study it is enough to focus on magnetic black holes, for which $F\tilde F =0$ and $F^2=2 B^2=2\frac{\e^2 Q_m^2}{r^4}$. The magnetic charge is denoted as $Q_m\equiv Q$ through the rest of the paper.

The black hole horizon satisfies $f(r_h) = 0$. From this, we find the relation between the mass $M$ and $r_h$:
\begin{equation}
    M = \frac{r_h}{\kappa^2} + \int^{r_h}_\infty dr r^2 \mathcal{L}_F\left[2B^2(r),0\right]\,, 
    \label{eq:BH_mass_gen}
\end{equation}
where $M=\frac{M_{\circ} }{4\pi}$. 
If there are two horizons, the condition for horizon degeneracy is
\begin{equation}
    \frac{d M}{dr_h} = \frac{1}{\kappa^2} + r_h^2 \mathcal{L}_F\left[2B^2(r_h),0\right] = 0\,.
    \label{eq:horizon_deg}
\end{equation}

\subsection{Euler-Heisenberg} \label{se:EH}

The Euler-Heisenberg effective action is a piece of the full electromagnetic one-loop effective action induced by a charged Dirac fermion of mass $m$ and fractional charge $q$. It is the part that neglects the higher-derivative corrections and 
encodes the full  field strength dependence. It contains thus precisely the information needed to  compute the extremal black hole in the strong-field regime in the presence of a charged particle. 

One can verify that the strong-field regime of the electric black hole 
matches the onset of decay via Schwinger effect, see e.g. \cite{Dunne:2004nc, Barbosa:2025uau}. For our study it is enough to focus on the magnetic black hole.

In the strong-field limit we have (see \cite{Dunne:2004nc} and references therein) 
\begin{align}
    \mathcal{L}_{\mathrm{EH}} 
   &  = -\frac{B^2}{2 \e^2} -\frac{ q^2B^2}{8\pi^2}\int^\infty_0\frac{ds}{s^2}\left(
    \coth s-\frac{1}{s}-\frac{s}{3}
    \right) e^{-\frac{m^2 s}{B q}} 
    \\ & \overset{B\gg \frac{m^2}{q}}{\approx}
     -\frac{B^2}{2}\left( \frac{1}{\e^2} - \frac{q^2}{24\pi^2}  \log\left(\frac{ q^2 B^2}{m^4} \right)\right)\,.
     \label{eq:L_HE}
     \end{align}
Using $B=\frac{eQ}{r^2}$ and \eqref{eq:f_gen}, we obtain the metric {factor} in the strong-field regime, 
\begin{equation}
    f(r) = 1 - \frac{\kappa^2 M}{r} + \frac{\kappa^2 Q^2}{2 r^2} 
\left(
1-\frac{q^2\e^2}{12\pi^2}\left( \log{\left(\frac{ \e q Q}{m^2 r^2} \right)}-2 \right) 
\right)\,.    
\end{equation}
The two horizon radii are corrected by the logarithmic term as 
\begin{align}
    r_{\pm} &= \frac{\kappa^2 M}{2} \pm \frac{\kappa}{2}\sqrt{\kappa^2 M^2 - 2 Q^2(1 - \delta_\pm) }\,,\quad\quad 
    \delta_\pm = \frac{q^2 e^2}{12 \pi^2}\left(-2 + \log{\left(\frac{\e qQ}{m^2 r_{\pm,0}^2} \right)} \right)\,,
\end{align}
with $r_{\pm,0}=r_\pm|_{\delta=0}$. 
The extremal black hole being defined by $r_+=r_-$, we find the extremal charge-to-mass ratio to be 
\be
    \bar Z_{\rm EH}\equiv \frac{\sqrt{2}Q}{\kappa M} =\frac{1}{\sqrt{1- \delta}} \approx 1 + \frac{q^2 \e^2}{24 \pi^2}\left(-2 + \log{\left(\frac{ \sqrt{2} \e q }{m^2 r_h\kappa} \right)} \right)\,,
\ee
where we have expanded in the loop factor and used that $Q\approx\frac{\sqrt{2}r_h}{\kappa}$ to simplify the logarithm. 

Having determined the extremal charge-to-mass ratio in the strong-field regime, we compute the variation in $M$ (or $r_h$) as 
\be M\frac{d}{d M} \bar Z_{\rm EH} \approx r_h\frac{d}{d r_h} \bar Z_{\rm EH}   = -\frac{q^2 e^2}{24\pi^2}\,.
\ee
It turns out that the variation is negative, therefore the charge growth bound \eqref{eq:WGC_sufficient_condition}
(see also \eqref{eq:charge_growth_bound}) is satisfied, and thus  any extremal black hole can decay. 

We conclude that  the black hole WGC is satisfied in Euler-Heisenberg electrodynamics.

\subsection{Dirac-Born-Infeld} \label{se:DBI}

The Dirac-Born-Infeld Lagrangian  \cite{Born:1934gh} is motivated by the low-energy effective action on D-branes \cite{Fradkin:1985qd}. In $d=4 $  it can be written as 
\begin{align}
    \mathcal{L}_{\mathrm{DBI}} &= \Lambda_{\mathrm{DBI}}^4 \left(1 - \sqrt{1 + \frac{ F^2}{2\e^2 \Lambda_{\mathrm{DBI}}^4} - \frac{(F  \Tilde{F})^2}{16 \e^4 \Lambda_{\mathrm{DBI}}^8}} \right)  \,.
    \label{eq:L_DBI}
\end{align}
We focus on the pure magnetic field, with $F^2=2B^2=\frac{2 e^2Q^2}{r^4}$. Applying the general formula \eqref{eq:f_gen} we obtain
the blackening factor
\begin{align}
    f(r) &= 1 - \frac{\kappa^2 M}{r} - \kappa^2 \Lambda_{\mathrm{DBI}}^4 r^2\left( \sqrt{1 + \frac{Q^{2}}{ \Lambda_{\mathrm{DBI}}^4 r^4}} - \frac{1}{3} + \frac{2}{3}\, {}_2 F_1\left(-\frac{3}{4}, \frac{1}{2}, \frac{1}{4}, -\frac{Q^{2}}{ \Lambda_{\mathrm{DBI}}^4 r^4} \right)  \right)\,. 
\end{align}
This matches the result in \cite{Abe:2023anf}. 

The study of the zeros of $f(r)$ shows that the black hole has either one or two  horizons depending whether $Q \kappa^2 \Lambda^2_{\rm DBI}$ is smaller or larger than one. This can be seen via the degeneracy condition \eqref{eq:horizon_deg}, which leads to 
\be
   r_h = \frac{\sqrt{Q^{2} \kappa^4 \Lambda_{\mathrm{DBI}}^4 - 1}}{\sqrt{2} \kappa \Lambda_{\mathrm{DBI}}^2}\,. 
   \label{eq:rh_DBI_deg}
\ee
Via the condition we can see that $r_h$ has two horizons only if $Q \kappa^2 \Lambda^2_{\rm DBI}>1$. Moreover, when  $Q \kappa^2 \Lambda^2_{\rm DBI}=1$ the extremal black hole has vanishing radius. 
For $Q \kappa^2 \Lambda^2_{\rm DBI}<1$, there is a single horizon. Extremality is reached for $r_h\to 0^+$, which is consistent with \eqref{eq:rh_DBI_deg}. 
This limit is well-defined, as explained in App. \ref{se:Growth_bound}.

We then identify the weak and strong-field regimes. 
The weak-field regime corresponds to $Q \kappa^2 \Lambda^2_{\rm DBI}>1$, as can be noted from \eqref{eq:rh_DBI_deg}. This case can be treated via the EFT expansion of \eqref{eq:L_DBI}. Positivity of the $F^4$ and $(F\tilde F)^2$ operators ensures that the charge growth bound \eqref{eq:WGC_sufficient_condition} is respected.

The strong-field regime corresponds to  $Q \kappa^2 \Lambda^2_{\rm DBI}<1$. Extremality in this limit is conveniently studied by observing the simplification 
\begin{align}
    \lim_{r \to 0^{+}} f(r) &= \lim_{r \to 0^{+}} \Bigg( 1 - \frac{\kappa^2 M}{r} - \frac{\kappa^2}{r}\int_r^\infty dr^\prime r^{\prime 2} \mathcal{L} \Bigg) \sim \frac{\kappa^2}{r} \Bigg(\frac{2 Q ^{3/2}\Lambda_{\text{DBI}}\Gamma{\left(\frac{1}{4} \right)}\Gamma{\left(\frac{5}{4} \right)}}{3 \sqrt{\pi}} - M \Bigg)\,
\end{align}
as noted in \cite{Abe:2023anf}. 
From this asymptotic result we conclude that an extremal black hole has mass  $\bar M=\bar M(Q)$ with
\be
 \bar M = \frac{2  Q^{3/2}\Lambda_{\text{DBI}}\Gamma{\left(\frac{1}{4} \right)}\Gamma{\left(\frac{5}{4} \right)}}{3 \sqrt{\pi}}\,. 
 \label{eq:DBI_M_ext}
\ee
 Values of $M$ below $ \bar M$ would feature a naked singularity. We invert the relation \eqref{eq:DBI_M_ext} to obtain the standard extremality curve as a function of $M$, $\bar Q=\bar Q(M)$. The $\bar Z(M)$ charge-to-mass ratio is
\be
\bar Z_{\rm DBI} = \frac{\sqrt{2}}{\kappa}\left( \frac{3 \sqrt{\pi}}{2  \Lambda_{\text{DBI}}\Gamma{\left(\frac{1}{4} \right)}\Gamma{\left(\frac{5}{4} \right)}}\right)^{\frac{2}{3}}  {M^{-\frac{1}{3}}} \,.
\ee
It follows that 
\be
\frac{d}{dM} \bar Z_{\rm DBI}  <0 \,.
\ee
Therefore the charge growth bound \eqref{eq:WGC_sufficient_condition}  is satisfied in the strong-field regime. Another convenient way to check the charge growth bound is to compute {$\frac{M}{Q}\frac{d Q}{d M}=\frac{2}{3}<1$}. 
Being a sufficient condition for extremal black hole decay, this ensures that any black hole can decay in the strong-field regime. 

We conclude that  the black hole WGC is satisfied in the
DBI model  in both weak and strong-field regimes --- even though the extremal black holes become tiny in the strong-field regime.

\subsection{ModMax}

\label{se:MM}

The ModMax extension of Maxwell's electrodynamics is defined by the Lagrangian \cite{Bandos:2020jsw,Kosyakov:2020wxv}
\be
{\cal L}_{\rm ModMax} = -\frac{\cosh\gamma}{4\e^2}F_{\mu\nu}F^{\mu\nu}+ \frac{\sinh\gamma}{4\e^2}\sqrt{(F_{\mu\nu}F^{\mu\nu})^2 + (F_{\mu\nu}\tilde F^{\mu\nu})^2 }
\ee
with $\gamma>0$. 
It has the interesting property of being both conformally invariant and invariant under duality rotations. Further generalizations have been proposed in \cite{Bandos:2020hgy,Kruglov:2021bhs}. For other developments, see
\cite{Sorokin:2021tge} and references therein.

The ${\cal L}_{\rm ModMax}$ Lagrangian does not have an EFT expansion, i.e. does not reproduce Maxwell in the small field limit, but rather for small $\gamma$. Therefore the usual black hole WGC criterion $\bar Z>1$, which  applies only in the EFT regime, could not apply here. Using the more general 
conditions \eqref{eq:WGC_sufficient_condition}, \eqref{eq:WGC_necessary_condition}  
is  mandatory to verify whether extremal black holes can decay in the ModMax model.

The charge-to-mass ratio of  charged black holes arising from ${\cal L}_{\rm ModMax}$ coupled to Einstein gravity is found in \cite{Flores-Alfonso:2020euz} (see also \cite{Sorokin:2021tge}) to be
\be
M=\frac{\sqrt{2}}{\kappa} \sqrt{Q_e^2+Q_m^2} e^{-\gamma}\,.
\ee
Extremal black holes in this model satisfy $\bar Z= e^{\gamma}$, such that 
\be
\frac{d\, \bar Z_{\rm ModMax}}{dM}=0\,.
\ee
Hence the sufficient condition {\eqref{eq:WGC_sufficient_condition}} is not satisfied. The necessary condition \eqref{eq:WGC_necessary_condition}, that requires the existence of a {discrete subset of masses where $\bar{Z}$ is decreasing, } 
is not satisfied either since $\bar{Z}_{\rm ModMax}$ is constant. This implies that the decay of extremal black holes is impossible, analogous to pure GR. 

We conclude that the ModMax model is not compatible with the black hole weak gravity conjecture. Whether ModMax generalizations like the one in \cite{Bandos:2020hgy, Sorokin:2021tge} can evade this outcome is an interesting direction for further investigation.

\section{Black Hole WGC and the $U(1)$ Renormalization Flow}

\label{se:BetaApplications}

\subsection{Beta Function Positivity}

\label{se:Beta_Positivity}

The quantum effective action taken in the strong-field limit has the form \cite{Weisskopf:1936hya,Matinyan:1976mp,Dittrich:2000zu,Pagels:1978dd,Fujikawa:1993xv,Grundberg:1994kk,Dunne:2004nc}
\be
\Gamma[F]\big|_{F^2\gg \Lambda^4_F}\approx \int d^4x \sqrt{g} \left(-\frac{1}{4g^2_{\rm eff}(t)}F_{\mu\nu}F^{\mu\nu}\right)\,,\quad\quad t=\frac{1}{4}\log\left( \frac{F^2}{\mu_0^4} \right) \,. 
\ee
This can be shown by applying standard renormalization group arguments to  $\Gamma[F, \mu^2 ]$, analogous to the application to the two-point function   $\Pi[q^2, \mu^2 ]$. In both cases, an external probe scale --- either $F$ or $q^2$ --- is varied to obtain the scaling behavior of the system. 
 This can also be shown via trace anomaly considerations \cite{Dunne:2004nc}.

 The $\mu_0$ scale is a reference scale at which a reference value of $g_{\rm eff}$ is defined --- and could be measured.  
The running of $g^2_{\rm eff}$ can be determined perturbatively. We define the beta function $\beta_g\equiv \mu\frac{d}{d \mu} g(\mu)$ and denote its leading term as $\beta_g=\beta^{(1)}g^3+\ldots$,  such that 
\be
\frac{1}{g^2_{\rm eff}(t)} = \frac{1}{g^2_0} -\frac{1}{2}\beta^{(1)} \log\left( \frac{F^2}{\mu_0^4}  \right)+{\rm higher~order~terms}
\label{eq:geff_running}
\ee
with $g_0\equiv g_{\rm eff}(0)$. For our argument it is enough to focus on the leading behavior shown in \eqref{eq:geff_running}.  
Notice that the strong-field behavior of the Euler-Heisenberg Lagrangian obtained in \eqref{eq:L_HE} is reproduced by setting $\beta^{(1)}$ to the one-loop beta functions $\beta^{(1)}_{{\rm Dirac}}= 4\beta^{(1)}_{{\rm scalar}}=\frac{1}{12\pi^2}$. 

The effect of the $U(1)$ renormalization flow on the extremal black hole solution in the strong-field regime is obtained by plugging the one-loop effective Lagrangian 
\be
{\cal L}_{\rm eff}[F]= -\frac{1}{4}\left(\frac{1}{g^2_0} -\frac{1}{2}\beta^{(1)} \log\left( \frac{F^2}{\mu_0^4}\right)\right)F^2
\ee
into the blackening factor \eqref{eq:f_gen}. 
Using $F^2=2\frac{g_0^2 Q^2}{r^2}$, the horizons radii are given by   
\be
  f(r) =  1 - \frac{\kappa^2 M}{r^2} + \frac{\kappa^2 Q^2}{2 r^2}\left(1 -\beta^{(1)} g_0^2 \log\left( \frac{ \sqrt{2} Q}{r^2\mu_0^2}\right)\right) = 0\,,
\label{eq:extemality_oneloop}
\ee
where we have already neglected the extra $-2$ term arising in the integration.

For the extremal black hole, using  $Q \approx \frac{\sqrt{2} r_h}{\kappa}$ in the logarithm,  the charge-to-mass ratio is found to be 
 \be
\bar Z = \frac{\sqrt{2}}{\kappa}\frac{Q}{M} \approx 1 + \beta^{(1)} \frac{g_0^2}{2} \log\left( \frac{ 2 g_0}{r_h\kappa\mu_0^2}\right)\,. 
 \ee
We  thus find the variation
\be
M\frac{d}{d M} \bar Z \approx r_h\frac{d}{d r_h} \bar Z   = -  \frac{g_0^2}{2}  \beta^{(1)}  \,. \label{eq:Z_beta_function}
\ee
This simple result explicitly shows  that the charge-to-mass ratio runs as a function of $r_h$ (or $M$). This may be viewed as a renormalization flow  in the space of black hole solutions. 
Equation \eqref{eq:Z_beta_function} makes clear that the notion of an absolute bound such as $\bar Z >1$ is ill-defined when one takes the $U(1)$ renormalization flow into account.

Let us inspect the two possible signs of $\beta^{(1)}$.  
If $\beta^{(1)}>0$, the sufficient condition \eqref{eq:WGC_sufficient_condition} is satisfied. In that case, all black holes can decay and the WGC is satisfied. 
If $\beta^{(1)}<0$, the $\bar Z$ function is strictly increasing, hence it does not feature any decreasing subset. Hence the necessary condition \eqref{eq:WGC_necessary_condition} is not satisfied, and the black hole WGC cannot be satisfied in that case. 
We conclude that the black hole WGC is satisfied if and only if \be \beta^{(1)}>0 \,. \label{eq:positivity}
\ee
The generalization to higher loop contributions is straightforward. 
It would be interesting to attempt a generalization of the argument at  non-perturbative level.

\subsection{Colored Black Holes}

\label{se:YM}

Non-Abelian gauge theory coupled to Einstein gravity features black hole solutions, see \cite{Volkov:1998cc} for a review.  
A subset of these are the so-called \textit{colored} or embedded-Abelian solutions, which were discovered long ago \cite{Bais:1975gu,Cho:1975uz,Wang:1975tu,Perry:1977wk,Kamata:1981nj}.
Here the term ``colored'' denotes specifically the Abelian solutions, not the neutral non-Abelian ones. 
The colored black holes correspond to the RN solution with electric  charge $Q$ and a unit magnetic charge.

The other type of solutions are neutral and non-Abelian \cite{Volkov:1989fi,Volkov:1990sva,Kuenzle:1990is,Bizon:1990sr, Baltsov:1991au,Kleihaus:1998qc,Kuenzle:1994ru,Kleihaus:1995tk, Kleihaus:1998qc}. These take the form of the RN metric in the large node limit (see \cite{Volkov:1998cc}), but with unit magnetic charge, and hence cannot be made extremal. Therefore the WGC argument cannot be applied to those, and they are not our focus. 
These solutions have been shown to be classically unstable \cite{Brodbeck:1994np,Brodbeck:1994vu}.

A common lore is that (Abelian)  colored black holes solutions  are  classically unstable \cite{Volkov:1998cc}.  Some instability for colored black holes in the presence of spontaneous symmetry breaking 
 via a Higgs field 
was found in \cite{Lee:1991qs}. But, to the best of our knowledge,  it seems that no general proof of instability has been  presented, including  for the case without spontaneous symmetry breaking.

We show that the black hole WGC provides a simple argument against the classical stability of colored black holes with unbroken gauge symmetry. 

Assuming a small enough number of flavors (e.g. $N_f \lesssim \frac{11}{2} N_c$ for Dirac fermions in the fundamental representation), the beta function of the Yang-Mills gauge theory is negative. 
This contradicts our general result from section \ref{se:Beta_Positivity} which states that the $U(1)$ beta function must be positive to be compatible with the black hole WGC.  In other words, from the decay of a colored black hole, we would find 
\be
\frac{d\,\bar Z_{\rm colored\,BH}}{d M} >0\,,
\ee
which would contradict the necessary condition \eqref{eq:WGC_necessary_condition}. Such a  contradiction is resolved if colored black holes are classically unstable.  
This argument does not apply if the gauge group is broken to its $U(1)$ subgroup, in which case the Abelian $U(1)$ running is recovered.

\subsection{Conformal Hidden Sector and AdS/CFT}

\label{se:CFT}

The $U(1)$ gauge theory experiences running  when it couples to a conformal sector. Let us investigate what the black hole WGC implies for this type of model. 

\subsubsection{4D Theory}

\label{eq:CFT}

The photon mixing to the conformal sector is described by\,\footnote{ This model is presented in rigorous form in formal AdS/CFT studies, where $A^\mu$ is a static source (see e.g. \cite{Witten:1998qj, Kap:lecture}), while it is often discussed only qualitatively when $A^\mu$ is dynamical, see e.g.  
\cite{ArkaniHamed:2000ds,Gherghetta:2010cj}. 
A similar analysis is done in \cite{Chaffey:2023xmz}   in the case of broken $U(1)$ symmetry.
}
\be
{\cal L} = -\frac{1}{4g_0^2}F_{\mu\nu}F^{\mu\nu} + a A_\mu {\cal J}^\mu[\varphi] +{\cal L}_{\rm CFT}[\varphi]
\label{eq:CFT_model}
\ee
where $\varphi$ denotes the fundamental degrees of freedom of the conformal sector and ${\cal J}^\mu[\varphi]$ is  the conserved $U(1)$ current of the CFT. The conservation equation together with conformal symmetries  constrain the conformal dimension of $\cal J$ to be exactly $\Delta_{\cal J}=3$ in four dimensions \cite{Minwalla:1997ka}. The two-point function has thus the form 
\be
\left\langle {\cal J}(0){\cal J}(x)\right\rangle= \frac{C_{\cal J}}{x^6}
\label{eq:2pt_CFT_pos}\,.
\ee
The normalization factor $C_{\cal J}$ is left undetermined by symmetries.

Integrating out exactly the CFT degrees of freedom gives the quantum effective action for the photon. The quadratic part can be written as  
\be
\Gamma[F] = - \frac{1}{4} \int d^4x  d^4y \sqrt{-g} F_{\mu\nu}(x)\Pi(x,y)F^{\mu\nu}(y) + \ldots = - \frac{1}{4} \int \frac{d^4p}{(2\pi)^4} \sqrt{-g} F_{\mu\nu}(p)\Pi(p)F^{\mu\nu}(-p) + \ldots
\ee
where the self-energy $\Pi(p)$ corresponds to the photon  inverse propagator dressed by insertions of the two-point function \eqref{eq:2pt_CFT_pos}. In Lorentzian momentum space, the two-point function reads  \cite{Costantino:2020vdu}
\be
\left\langle {\cal J}(p){\cal J}(-p)\right\rangle= -i\frac{\pi^2}{24} C_{\cal J} \left(p^2 \log\left(\frac{p^2}{\mu^2}\right) +{\rm cst}\right)\,.
\label{eq:2pt_CFT}
\ee
where $p^2>0$ corresponds to spacelike momentum. 
The arbitrary mass scale $\mu$ and the constant terms in \eqref{eq:2pt_CFT}  are ultimately absorbed into the definition of $g_0$. 

By direct calculation of the dressed photon propagator 
\be
\frac{ig_{\mu\nu}}{p^2}+\frac{ig_{\mu\nu}}{p^2} ia \left\langle {\cal J}(p){\cal J}(-p)\right\rangle ia \frac{ig_{\mu\nu}}{p^2} +\ldots 
\ee
we obtain the exact result
\be
\Pi(p^2) = \frac{1}{g_0^2}+\frac{a^2\pi^2 }{24}C_{\cal J} \log\left(\frac{p^2}{\mu^2} \right)\,. 
\label{eq:2pt}
\ee
We can read from the self-energy \eqref{eq:2pt} the  renormalization flow of the $g$ coupling induced by the mixing to the CFT: 
\be
\beta_g = -g^3 \frac{a^2\pi^2}{24} C_{\cal J}\,. 
\label{eq:beta_g_CFT}
\ee

We emphasize that even though this beta function  \eqref{eq:beta_g_CFT} is obtained
via momentum dependence of the two-point function,  \eqref{eq:2pt}, it describes a fundamental property of the theory. This $U(1)$ renormalization flow readily applies to the strong-field behavior by virtue of the standard approach described in section \ref{se:Beta_Positivity}. 
We thus use our positivity result from section \ref{se:Beta_Positivity} to conclude that the model defined by \eqref{eq:CFT_model} is compatible with the black hole WGC if 
\be
a^2 C_{\cal J}<0\,. \label{eq:CJ_bound}
\ee

The model \eqref{eq:CFT_model} is often written schematically in the literature, see e.g. \cite{ArkaniHamed:2000ds}, in which case the bound \eqref{eq:CJ_bound}  cannot be taken into account. The bound constrains more detailed models  of conformal hidden sectors such as  \cite{Redi:2021ipn,Chiu:2022bni}, that  feature \eqref{eq:CFT_model} as interactions between the elementary sector and the hidden CFT.

\subsubsection{Holographic Theory}

\label{eq:AdS}

The 4d CFT model of section \ref{se:CFT}  taken in the large $N$ limit is equivalently described by a 5d theory with a flat 3-brane. The quantum effective action supported on the brane reproduces the structure of \eqref{eq:CFT_model} 
as a manifestation of the   AdS/CFT correspondence. The computation of the $\Pi$ self energy has been performed in a number of references to various degrees of accuracy, see for example \cite{Pomarol:2000hp, ArkaniHamed:2000ds,Contino:2002kc,Randall:2001gb,Goldberger:2002hb,Friedland:2009zg,Fichet:2019owx}.\,\footnote{
Holographic realizations of conformal hidden sectors have been proposed in \cite{vonHarling:2012sz,McDonald:2012nc,McDonald:2010fe,McDonald:2010iq,Brax:2019koq,Chaffey:2021tmj}. 
 The running of the  $U(1)$ gauge coupling as a phenomenological signature of the AdS braneworld has been discussed in \cite{Fichet:2019owx}. 
} 
We get
\be
\Pi^{\rm AdS}(p^2) = \frac{1}{g_0^2}-\frac{L}{g_5^2 } \log\left( \frac{p^2}{\mu^2} \right)\,,
\label{eq:2pt_AdS}
\ee
where $L$ is the AdS radius and $g_5$ is the bulk gauge coupling. The $g_0$  encapsulates constant contributions including a brane-kinetic term.
In the presence of an infrared brane, the $\mu$ scale can be taken as the IR brane scale in which case $g_0$ corresponds to the 4d gauge coupling of the low-energy gauge mode.\,\footnote{
In the presence of the IR brane, the form \eqref{eq:2pt_AdS} holds for $p^2\gg \mu^2$, 
for which the IR region of the AdS bulk becomes opaque to propagation \cite{ArkaniHamed:2000ds, Fichet:2019hkg, Costantino:2020vdu}, so that the effect of the IR brane vanishes from the UV correlators.  
}

We read from \eqref{eq:2pt_AdS} that the beta function of  $g$ is positive: 
\be
\beta_g^{\rm AdS}= \frac{L}{g_5^2}>0\,. 
\ee
Therefore the $U(1)$ renormalization flow arising in holographic models satisfies the charge growth bound, i.e. it allows extremal black holes to decay.  
In other words, the black hole WGC is consistent with AdS/CFT.

\section{Summary }
\label{se:conclusion}

In this note we study the  interplay between non-spinning extremal black holes, strong electromagnetic fields and  the WGC. 

We first observe that  the electromagnetic field near the horizon of a charged black hole can either be in the weak-field regime, i.e. described by the Maxwell EFT, or in the strong-field regime, which depends on the UV completion of the Maxwell sector. 
We point out that for extremal black holes,  the existence of the strong-field regime is ensured by an EFT positivity bound which is derived from applying the black hole WGC in the weak-field regime. Therefore,  sufficiently small extremal black holes  generically probe the strong-field UV completion of the Maxwell sector.
This is summarized in Fig.\,\ref{fig:Sketch_Large_Field}. 
 Conveniently, whenever the positivity bound is not saturated,  the extremal black hole is dominated by Maxwell corrections, which are much easier to compute than gravitational ones.

We then revisit the black hole WGC --- the conjecture that extremal black holes of any size can decay --- to derive  conditions valid beyond the weak-field regime,  for any gravitational field theory in asymptotically flat space. 
We derive a  sufficient condition which is continuous and a necessary condition which is discrete. They can be summarized as
\be
 \frac{d \bar Z }{dM}<0~~\forall M~~ \Rightarrow~~ {\rm Extremal~BH~decay} ~~\Rightarrow ~~ \exists A=\{M_n\} {\rm ~such~that~}  Z|_A~{\rm decreases}\,.
\ee
 The sufficient condition is equivalently stated as a bound on charge growth $\frac{d \bar Q}{d M} <\frac{\bar Q}{M}$, for all $M$. 
The trick to derive this condition is to obtain that $\bar Q$ is subadditive.

We apply our conditions for extremal black hole decay to a few strong-field UV completions of the pure Maxwell sector. 
Focusing on magnetic black holes for simplicity, we find that both Euler-Heisenberg and  DBI effective actions satisfy the sufficient condition, and are thus compatible with the black hole WGC. 
In contrast, the ModMax model does \textit{not} satisfy the necessary condition, hence extremal black holes cannot decay in this model, analogous to the pure GR case. 

We show that the renormalization flow of the $U(1)$ gauge coupling implies that charge-to-mass ratio of the black hole solutions varies logarithmically with $r_h$. This can be viewed as a renormalization flow in the space of black hole solutions, with the renormalization scale identified as  $\frac{1}{r_h}$.
The computation is done here at  the perturbative level --- it would be interesting to attempt an analogous one at the nonperturbative level. It turns out that  the
conditions for extremal black hole decay
 constrain the sign of the $\log(r_h)$ dependence of $\bar Z$, implying that the $U(1)$ beta function must be positive to be consistent with the black hole WGC. 

Beta function positivity  provides an independent argument against the existence of colored black holes, since for such solutions the beta function is non-Abelian and can thus be negative. 
This matches the common lore that colored black hole solutions are classically unstable \cite{Volkov:1998cc}, in which case the decay arguments do not apply. 
Beta function positivity also constrains the sign of a combination of parameters in the EFT that describes a $U(1)$ gauge field coupled to a (nearly) conformal sector with $U(1)$ charge. The holographic  AdS realization of such models feature a classical beta function that is always positive,  thus ensuring compatibility of the black hole WGC with the AdS/CFT correspondence.

\begin{acknowledgments}

We thank  Dmitri Vassilievich and Philippe Brax for useful discussions. 
SF thanks the IPhT/CEA-Saclay for funding  a visit during which this work was initiated. 
This work was supported in part by the S\~ao Paulo Research Foundation (FAPESP), grant 2021/10128-0. The work of LS and SB was supported by grant 2023/11293-0 and 2025/05571-3 of FAPESP. The work of SB has been supported in part by grant 001 of CAPES.

\end{acknowledgments}

\appendix

\section{Conditions for the Black Hole WGC  }

We derive  conditions, either necessary or sufficient, for the decay of any black hole 
in a gravitational theory with a generic Maxwell sector in asymptotically flat spacetime. 

\label{se:Growth_bound}

\subsection{On the Existence of the Extremality Curve}

The corrections to a sufficiently charged black hole are dominated by the Maxwell contributions, as discussed in section \ref{se:Maxwell_dom}. Neglecting the higher curvature corrections to gravity,  the mass of the black hole  with charge $Q$ in asymptotically flat spacetime satisfies a relation of the form\,\footnote{See \eqref{eq:BH_mass_gen} for the explicit expression in the magnetic case.}
\be
M=\frac{r_h}{\kappa^2} + M_F(Q,r_h)\,. \label{eq:mass_components}
\ee
The $\frac{r_h}{\kappa^2}$ contribution can be interpreted as the bare mass of the black hole, i.e. the total energy trapped inside the event horizon. The $M_F$ term is the total energy of the electromagnetic fields dressing the black hole. This energy is the same as for any other  electrically or magnetically charged spherical object. 
Thus even though the ${\cal L}_F$ Lagrangian is mostly unspecified, we  assume that $M_F$ is positive in all configurations.  

One implication is that $M$ is nonzero for any $Q$ and $r_h$. Thus for a given $Q$, the minimal value of $M$  is nonzero. Hence the charge-to-mass ratio $Z$ is bounded from above. In other words, the extremality curve $Z(M)$ always exists.

Even if the radius becomes tiny, the black hole still has finite $Z$, similarly to an elementary particle. 
The fundamental difference between a sub-Planckian elementary particle and a black hole is that the latter must be subextremal. Subextremality  highly constrains the decay kinematics when the only available final states are other black holes. This is the configuration considered in this work, which is  useful to constrain black hole extremality.

\subsection{Single Charge}
\label{se:singlecharge}

We consider a single $U(1)$ gauge group.  We denote the  charge and mass of the black holes by $Q$ and $M$, respectively, and define the charge-to-mass ratio 
\be
Z=\frac{\sqrt{2}}{\kappa}\frac{Q}{M}\,. 
\ee
The upper bound on $Z$, that we denote $Z|_{\rm extremal} \equiv \bar Z $, corresponds to extremal black holes.  It is convenient to think of it as a function of the black hole mass, $\bar Z = \bar Z (M)$. 
This defines the black hole extremality curve. Any black hole must satisfy $Z\leq \bar Z$ for any $M$, i.e. lie below the extremality curve.

The process of our focus is the decay of a black hole into smaller ones: BH$_0\to \sum_i{\rm BH_i}$ with $i=1,2,\ldots$ 
Charge is conserved while some mass is dissipated into gravitational waves. Hence the offspring black holes  satisfy
\be
Q_0=\sum_i Q_i\,,\quad\quad  M_0 > \sum_i M_i\,.
\ee
We introduce the mass fraction $\sigma_i=\frac{M_i}{M_0}$
such that 
\be
  \sum_i \sigma_i < 1  \,.
\label{eq:sum_Z}
\ee
The borderline case $ \sum_i \sigma_i Z_i = 1 $ would require no  gravitational wave emission in the decay process, which can  be considered as impossible. Note that in QFT such a kinematic configuration would have probability zero.

Using the above definitions, a black hole is allowed to decay if there exists a set of  $\sigma_i$, $Z_i$ such that 
\be Z_0 = \sum_i \sigma_i Z_i\,. 
\label{eq:decay_condition}
\ee

\subsubsection{General Relativity}\label{sec:gr}
Let us first review what happens in general relativity. In GR, extremal black holes of any size  satisfy  $\bar Z(M) = 1$ for any $M$. 
The decay products of an extremal black hole BH$_0$ ($\bar Z_0=1$) satisfy 
\be
1=\sum_i \sigma_i Z_i \,,
\ee
{with $Z_i\leq \bar{Z}(M_i)$.} 
Rearranging as 
\be
\sum_i \sigma_i (Z_i-1) =1- \sum_i \sigma_i  \,  \label{eq:sum_1charge}
\ee
and taking into account the mass inequality $\sum_i \sigma_i \leq 1$, we see that the l.h.s of \eqref{eq:sum_1charge} is $\leq 0$ while the r.h.s is $\geq 0$. It follows that \eqref{eq:sum_1charge} can hold only if all offspring black holes are extremal ($Z_i=1$) \textit{and} if there is no mass dissipation ($\sum_i \sigma_i = 1$). This borderline line case with no mass dissipation is excluded as discussed earlier in   section \ref{se:singlecharge}.

The logical conclusion is that, if extremal black hole must decay, then a deviation to the GR relation $\bar Z =1$ must occur. In the following we present two conditions on the extremality curve, one  sufficient and one necessary,  governing the decay of extremal black holes.

\subsubsection{A sufficient condition}

\label{se:proof_sufficient_singlecharge}

Here we  show that the monotonicity condition  {$\frac{d\Bar{Z}}{dM} < 0$} on the extremality curve   constitutes a sufficient condition for the decay of extremal black holes.

First, observe that the strict monotonicity condition $\bar{Z}^\prime < 0$ implies the strict subadditivity of the extremal charge function $\bar{Q}$.\footnote{Indeed, for any $M_1, M_2 > 0$, since $\bar{Z}$ is strictly decreasing, we have $\bar{Z}(M_1), \bar{Z}(M_2)  > \bar{Z}(M_1+M_2)$. Therefore, 
\be \bar{Q}(M_1+M_2) = (M_1 + M_2)\bar{Z}(M_1+M_2) < M_1 \bar{Z}(M_1) + M_2\bar{Z}(M_2) = \bar{Q}(M_1) + \bar{Q}(M_2)\,.\ee}
Hence, given arbitrary positive masses $\tilde{M}_1, \tilde{M}_2$ such that $ \tilde{M}_1 + \tilde{M}_2 = M_0$, by subadditivity of the function $\bar{Q}(M) = M \bar{Z}(M)$ we have
\be
 M_0\bar{Z}(M_0) < \tilde{M}_1\bar{Z}(\tilde{M}_1) + \tilde{M}_2\bar{Z}(\tilde{M}_2)\,.
\ee
Since the inequality is strict, by continuity there must exist positive masses $M_i < \tilde{M}_i$ such that
\be
 M_0\bar{Z}(M_0) < M_1\bar{Z}(\tilde{M}_1) + M_2\bar{Z}(\tilde{M}_2)\,. 
\ee
By defining the mass-fractions $\sigma_i = \frac{M_i}{M_0}$ and using the decreasing property $\bar{Z}(\Tilde{M}_i) < \bar{Z}(M_i)$ we can write
\be
\bar{Z}(M_0)<\sigma_1 \bar{Z}(M_1)+\sigma_2 \bar{Z}(M_2)\,.
\ee
It is therefore possible to choose black holes with $Z_i \leq \bar{Z}(M_i)$ such that the equality \eqref{eq:decay_condition} be satisfied. 

Note that this argument applies to any partition $\Tilde{M}_1, \dots, \Tilde{M}_n$ of the parent mass $M_0$. Summarizing, we arrive at the relations 
\be
\bar{Z}(M_0)= \sum_{i=1}^n \sigma_i Z_i,  \quad\quad \sum_{i=1}^n \sigma_i < 1,  \quad\quad Z_i < \bar{Z}(M_i)\,.
\ee
This is precisely the conditions that define decay of extremal  black holes, as discussed earlier in section  \ref{se:singlecharge}.

We conclude that the strict monotonicity condition on $\bar{Z}$ implies that extremal black holes can decay, proving \eqref{eq:WGC_sufficient_condition}. 
This condition is also conveniently expressed in terms of the black hole extremal charge  $\bar Q = \frac{\kappa M}{\sqrt{2}}\bar Z$, with   $\bar Q=\bar Q(M)$.  
    In summary, our sufficient condition for the black hole WGC is  
     \be
\hspace{3cm}  \frac{d \bar Z}{d M}<0\quad \quad  \left[ {i.e.}~~\frac{d\, \bar Q}{d\, M} <\frac{\bar Q}{M}\right]\, \hspace{1cm} { (Charge~Growth~Bound ) }
   \label{eq:charge_growth_bound}
   \ee
In any EFT in which the extremality curve satisfies  the bound \eqref{eq:charge_growth_bound} for all $M$, all black holes can decay.

\subsubsection{A necessary  condition}

\label{se:proof_necessary_singlecharge}

Here we  show that the condition of extremal black hole decay implies a {discrete} decrease condition on the extremality curve.

An extremal black hole BH$_0$ of mass $M_0$ and charge-to-mass ratio $\bar{Z}_0$ can decay if there exist black holes of masses $M^{(0)}_1, \dots, M^{(0)}_n$ and charge-to-mass ratios $Z^{(0)}_i$ satisfying
\be \bar{Z}_0 = \sum_{i=1}^{n} \sigma_i Z^{(0)}_{i}\,, \quad  \sum_{i=1}^{n} \sigma_i < 1\,. \label{eq:ebh_decay}\ee
Let $Z^{(0)} = \operatorname{max}\{Z^{(0)}_1 , \dots , Z^{(0)}_n\}$ be the maximum charge-to-mass ratio among the decay products and $M^{(0)}$ the corresponding mass. 
Since  $\sum \sigma_i < 1$,  and $Z_i^{(0)} \leq \bar{Z}_i^{(0)}$
with $\bar{Z}_i^{(0)}\equiv \bar{Z}(M_i^{(0)})$, we have
\bea
\bar{Z}_0 &=&  \sum_{i=1}^n \sigma_{i} Z^{(0)}_{i}   \leq Z^{(0)}\sum_{i=1}^n \sigma_{i} < Z^{(0)}   \leq \bar{Z}^{(0)}\,,
\eea
such that $\bar{Z}_0<\bar{Z}^{(0)}$, with $M_0 > M^{(0)}$ .

This process can be applied iteratively. {Assume we have defined a set of masses $A_k = \{M_0, M_1, \dots, M_{k-1}\}$, ordered as $M_0 > M_1 > \dots > M_{k-1}$, where the associated charge-to-mass ratios satisfy $\bar{Z}_0<\bar{Z}_1<\dots < \bar{Z}_{k-1}$.} The next element, $M_{k} \equiv M^{(k-1)}$, is defined by considering the decay of the extremal black hole $\operatorname{BH}_{k-1}$, with mass $M_{k-1}$ and charge-to-mass ratio $\bar{Z}_{k-1}$. In this manner, we inductively construct a discrete set $A = \bigcup_{k \in \mathbb{N}} A_k$ for which the function $\bar{Z}|_{A}$ is strictly decreasing.

In summary, the condition of extremal black hole decay implies there {must exist a} discrete subset of masses $A$ over which $\bar{Z}|_{A}$ is strictly decreasing, which proves \eqref{eq:WGC_necessary_condition}.\footnote{It is possible to show that this discrete necessary condition becomes {a} sufficient {condition} when additionally requiring subadditivity of $\bar{Q}$ with respect to finite sums in $A$, i.e.
\be
\bar{Q}(M_1 + \dots + M_n) < \bar{Q}(M_1) + \dots + \bar{Q}(M_n), \quad  \forall n \in \mathbb{N}_{\geq 2}, \forall M_1, \dots , M_n \in A\,.
\ee}

\subsection{Multiple Charges}

Here we  extend our decay conditions to  black holes charged under a  product of gauge groups  $U(1)\times U(1)' \times \ldots $ The corresponding charges  are then represented as {a vector $  {\bm Q} \equiv (Q, Q',\ldots) $}, and the charge-to-mass ratio {vector} is 
\be
 \Zvec=\frac{\sqrt{2}}{\kappa}\frac{\bm Q}{M}\,. 
\ee
Both electric and magnetic charges are  included into the $\bm Q$ vector as $\bm Q=({\bm Q}_{\rm e}, {\bm Q}_{\rm m}) $. 
A black hole is extremal when $|\Zvec|=\bar Z $. 

For multiple charges, extremal black holes are described
by an extremality surface depending on $M$ and on the direction in the space of charge-to-mass ratios that we denote by $\hat{\bm n}$. 
It is convenient to think of $\bar Z=\bar Z(\hat{\bm n},M)$ as the modulus of the extremality vector of charge-to-mass ratios $\bar \Zvec=\bar \Zvec(\hat{\bm n},M)$ (or  $\bar {\bm Q} = \frac{\kappa M}{\sqrt{2}}\bar \Zvec$ for charges) that describes the black hole extremality surface.  
In practice we will not need to know the individual components of $\bar \Zvec$, only its modulus. Figure \ref{fig:surface} illustrates the extremality surface together with the key properties underlying the generalization of the sufficient and necessary conditions for multicharged black hole decay.

The offspring black holes satisfy 
\be
 \Zvec_0 = \sum_i \sigma_i \Zvec_i \,.
\label{eq:sum_Z_vec}
\ee
The situation is mathematically analogous to the decay into superextremal particles studied in \cite{Cheung:2014vva} (see also \cite{Jones:2019nev}), up to the key difference that here the decay products are subextremal. 
The r.h.s of \eqref{eq:sum_Z_vec} defines the convex hull of the $\Zvec_i $ vectors. 
A black hole can decay if its charge-to-mass vector ${\bm Z}_0$ ends inside the convex hull.

\begin{figure}
    \centering
    \includegraphics[width=0.5\linewidth]{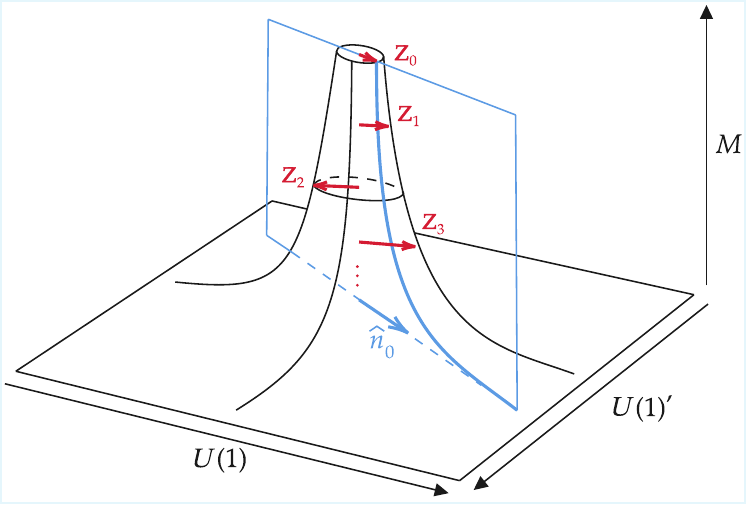}
    \caption{     \label{fig:surface}
An example of extremality surface for  black holes charged under $U(1) \times U(1)^\prime$.  If the surface satisfies the monotonicity condition $\frac{\partial \bar{Z}(\hat{\bm n},M)}{\partial M} <0 $, then the direction $\hat{n}_0 = \Zvec_0/|\Zvec_0|$ of any extremal black hole BH$_0$ defines a curve along which the norm of the extremal charge-to-mass ratio vectors decreases with the mass $M$ (blue), which is a property used in the proof of the sufficient condition \eqref{eq:WGC_sufficient_condition}. On the other hand, the decay of extremal black holes allows the construction of a sequence of vectors on the extremality surface whose norm increases as the corresponding mass decreases (red), which is used in the proof of the necessary condition \eqref{eq:WGC_necessary_condition}.} 
    \label{fig:surface}
\end{figure}

\subsubsection{General Relativity}

In GR, extremal black holes of any mass and  any direction satisfy $\bar Z=1$. The decay products of an extremal black hole BH$_0$ ($\bar Z_0=1$) satisfy 
\be
1=\left|\sum_i \sigma_i \Zvec_i \right| \,. 
\ee
Squaring and rearranging as 
\be
\sum_i \sigma^2_i (|\Zvec_i|^2-1) + 2 \sum_{i>j} \sigma_i \sigma_j (\Zvec_i \cdot \Zvec_j-1) =1- \left(\sum_i \sigma_i\right)^2 \,  \label{eq:sum_Ncharge}
\ee
we see that the l.h.s  is $\leq 0$ while the r.h.s is $\geq 0$. Hence, like in the one-charge case, \eqref{eq:sum_Ncharge} holds only if there is no mass dissipation and all offspring black holes are both extremal and aligned (i.e. $\Zvec_i=\Zvec_j$).   

In geometric terms, the convex hull of the $\Zvec_i$ vectors is inside the unit ball everywhere.  The intersection of the convex hull with the boundary is a single point (and its charge conjugate), corresponding to   the alignment limit in which case the convex hull degenerates to a segment of length two.   

\subsubsection{A sufficient condition}

Here we show that the monotonicity condition,  generalized to the case of multiple charges as
$\frac{\partial \bar{Z}_{\hat{\bm n}}}{\partial M} <0 $
for {\textit{all}} directions $\hat{\bm n}$, {where} $\bar{Z}_{\hat{\bm n}}(M) \equiv \bar{Z}(\hat{\bm n}, M)$, constitutes a sufficient condition for the decay of extremal black holes carrying multiple charges.

{Let BH$_0$ be an extremal black hole with charge-to-mass ratio vector $\Zvec_0$.} 
We observe that, by considering the function $\bar{Z} \equiv \bar{Z}_{\hat{\bm{n}}_0}$ with fixed direction  $\hat{\bm{n}}_0 = \Zvec_0/|\Zvec_0|$ , the problem reduces to the one-charge case, with all black holes configurations living below the curve   $\bar{Z}_{\hat{\bm{n}}_0}(M)$.

Applying the proof of \eqref{eq:WGC_sufficient_condition}, 
we conclude that there exist black holes BH$_i$ with masses $M_i$ and 
charge-to-mass ratios $\Zvec_i = Z_i\hat{\bm{n}}_0$, with $Z_i \leq \bar{Z}(\hat{\bm n}_0, M_i)$,
 such that  $\sum \sigma_i <1$ and $\bar{Z}(M_0) = \sum \sigma_i Z_i$. 
{Therefore,}
\be
\sum_i \sigma_i \Zvec_i  =  \left(\sum_i \sigma_i Z_i \right) \hat{\bm{n}}_0 = \bar{Z}(M_0)\hat{\bm{n}}_0 = \Zvec_0\,,
\ee
 so that $\operatorname{BH}_0$ decays into $\sum \operatorname{BH}_i$ along the $\hat {\bm n}_0$ direction.

It follows that requiring $\bar{Z}_{\hat{\bm{n}}}^{\prime}<0$ for \textit{all} $ \hat{\bm{n}}$ implies that any multicharged extremal black hole can decay.
This condition is also conveniently expressed in terms of the black hole extremal charge vector $\bar{\bm{Q}} = \frac{\kappa M}{\sqrt{2}} \bar{\Zvec}$, with $\bar{\bm{Q}} = \bar{\bm{Q}}(\hat{\bm{n}},M)$.  In summary, our sufficient condition for the multicharged black hole WGC is
   \be
 \hspace{1cm}  \frac{\partial \bar Z(\hat{\bm{n}},M)}{\partial M}<0 \quad \quad \left[i.e.~~\frac{\bar {\bm Q}}{\bar Q} \cdot \frac{  \partial \bar {\bm Q}}{\partial M} <\frac{{\bar Q}}{M}\right] \, \hspace{1cm} { (Multicharge~Growth~Bound ) }
    \label{eq:multicharge_growth_bound}
    \ee
In any EFT in which the extremality surface satisfies the bound \eqref{eq:multicharge_growth_bound} for all $M$ and all directions $\hat{\bm{n}}$, all black holes can decay.

\subsubsection{A necessary  condition}

Here we  show that the condition of extremal black hole decay implies a discrete condition on the extremality surface.

An extremal black hole BH$_0$ of mass $M_0$ and charge-to-mass ratio vector $\Zvec_0$ can decay if there exist black holes of masses $M^{(0)}_1, \dots, M^{(0)}_n$ and charge-to-mass ratio vectors $\Zvec^{(0)}_i$ satisfying
\be \Zvec_0 = \sum_{i=1}^{n} \sigma_i \Zvec^{(0)}_{i}\,, \quad  \sum_{i=1}^{n} \sigma_i < 1\,. \label{eq:ebh_decay}\ee
Let $Z^{(0)} = \operatorname{max}\{|\Zvec^{(0)}_1| , \dots , |\Zvec^{(0)}_n|\}$ be the maximum charge-to-mass ratio vector {norm} among the decay products, with $\Zvec^{(0)}$ being the corresponding charge-to-mass ratio vector and $M^{(0)}$ the corresponding mass. 
Since  $\sum \sigma_i < 1$ and $|\Zvec^{(0)}_i| \leq \bar{Z}^{(0)}_i$, with $\bar{Z}^{(0)}_i \equiv \bar{Z}(\hat{\bm n}^{(0)}_i, M^{(0)})$, by triangle inequality we have
\bea
\bar{Z}_0 = |\Zvec_0| = \left|\sum_i \sigma_i \Zvec^{(0)}_i\right|\leq \sum_i \sigma_i |\Zvec^{(0)}_i| \leq Z^{(0)} \sum_i \sigma_i  < Z^{(0)} \leq \bar{Z}^{(0)}\,,
\eea
therefore the directions $\hat{\bm n}_0$ and $\hat{\bm n}^{(0)}$ are such that $\bar{Z}_0<\bar{Z}^{(0)}$, with $M_0 > M^{(0)}$. Analogous to the one-charge case, this process can be applied iteratively to construct a discrete set of pairs $A = \{(\hat{\bm n}_0, M_0), (\hat{\bm n}_1, M_1), \dots \}$ by defining $(\hat{\bm n}_k,M_k) \equiv (\hat{\bm n}^{(k-1)},M^{(k-1)})$, for which the function $\bar{Z}|_{A}$ is strictly decreasing with respect to the mass.

In summary, the condition of extremal black hole decay for multiple charges implies there {must exist} a discrete subset of direction-mass pairs $A$ over which $\bar{Z}|_A$ is strictly decreasing {with respect to the mass}.\,\footnote{Note that in the one-charge case treated in \ref{se:proof_necessary_singlecharge}, the direction is {unique} and thus  need not be specified. }

\label{se:Decay_proof}

 \subsubsection*{Example:  $U(1)\times U(1)'$}

 \begin{figure}[t]
     \centering
    \includegraphics[width=0.45\linewidth,trim={8cm 5cm 4cm 3cm},clip]{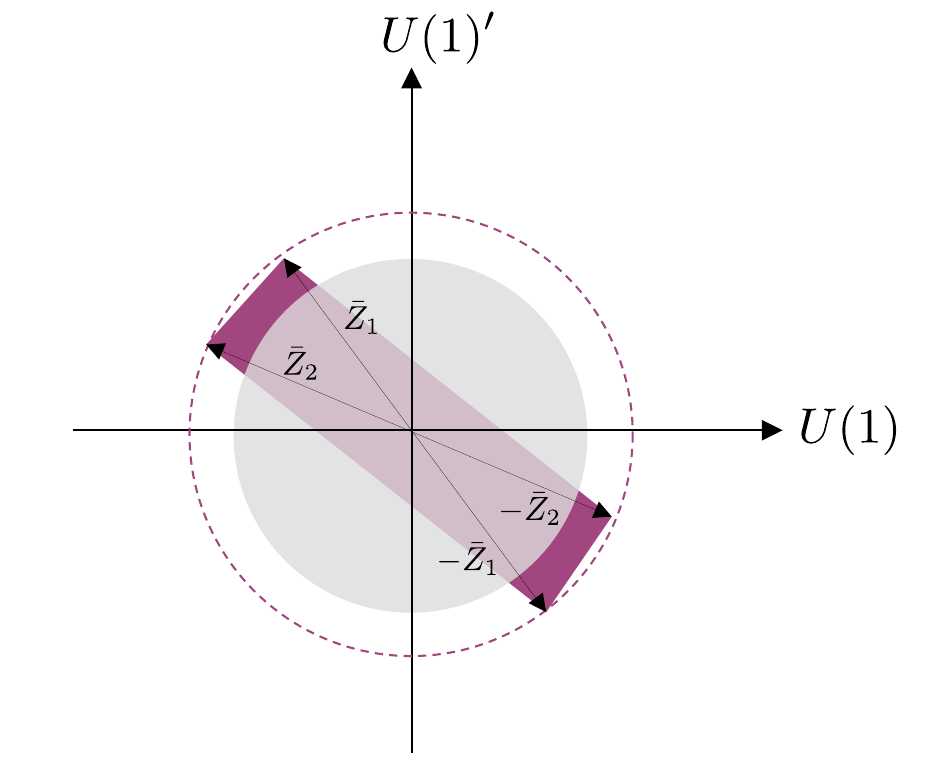}
        \includegraphics[width=0.45\linewidth,trim={8cm 5cm 4cm 3cm},clip]{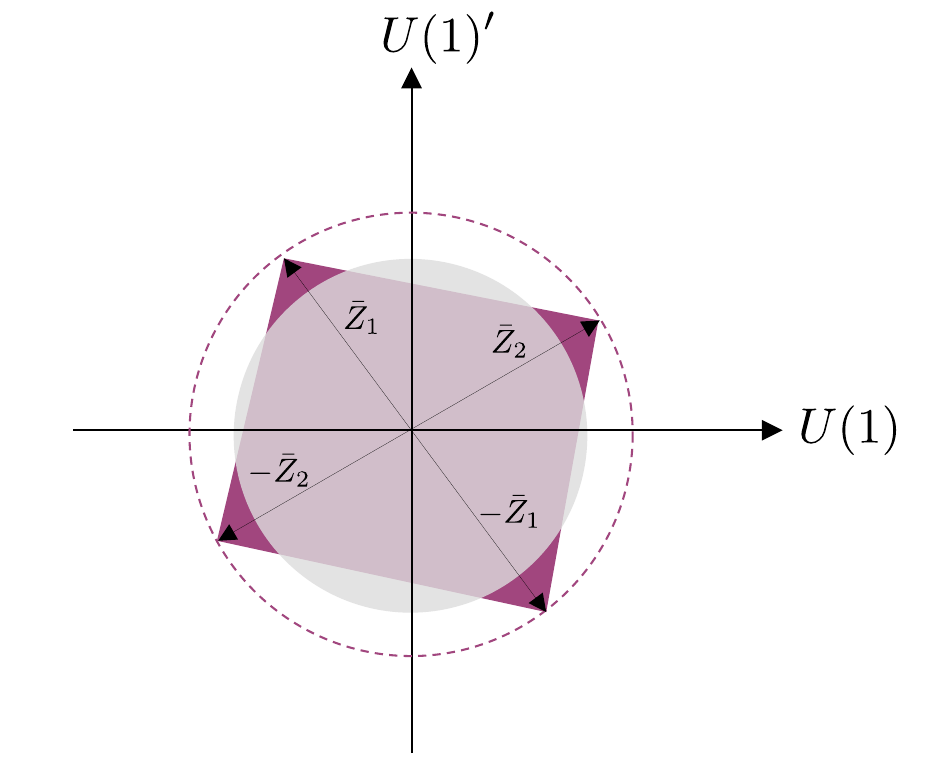}
\caption{     \label{fig:CH} 
Kinematic configurations for a black hole BH$_0$  decaying into two black holes BH$_{1,2}$ charged under  $U(1)\times U(1)'$. The  charge-to-mass  vector 
${\bm Z}_0$ of  BH$_0$ spans the gray volume, whose boundary corresponds to the extremality surface $Z_0=\bar Z_0$.
The charge-to-mass vector  of BH$_{1,2}$ is encoded in ${\bm Z}_1$, ${\bm Z}_2$. For given ${\bm Z}_{1,2}$,  the allowed configurations for the decay of BH$_0$  are given by the intersection of the gray volume with the  convex hull of  ${\bm Z}_1$, ${\bm Z}_2$, shown in purple. 
The extremality surfaces of BH$_1$ and BH$_2$ are assumed to be equal for simplicity $Z_{1,2}=\bar Z_{1,2}$, which is represented by the dashed line. 
Left configuration:  ${\bm Z}_1$ and ${\bm Z}_2$ are approximately colinear,  the decay of extremal BH$_0$ can be symmetric. 
Right configuration: ${\bm Z}_1$ and ${\bm Z}_2$ are not  colinear, the decay of extremal BH$_0$  is asymmetric.
 } 
 \end{figure}

As an example, we analyze the kinematic configurations 
 for  a black hole   decaying into two  black holes charged under two Abelian groups $U(1)$ and $U(1)'$. This is  represented in Fig.\,\ref{fig:CH}. 

For given charge-to-mass ratio vectors ${\bm Z}_1$, ${\bm Z}_2$, the parent black hole can decay if its ${\bm Z}_0$ vector ends inside the convex hull of ${\bm Z}_1$, ${\bm Z}_2$, as represented in  Fig.\,\ref{fig:CH}. Hence the parent extremal black hole can decay into configurations given by the intersection of the extremality surface with the convex hull. In GR this intersection would be a mere point, as shown earlier in this section. In contrast, if $\bar Z_0<\bar Z_{1,2}$ in all directions, as required by the charge growth bound \eqref{eq:multicharge_growth_bound}, the intersection of the extremality surface with the convex hull has nonzero dimension, such that decay is kinematically allowed.

It is worth noticing that two qualitatively different cases appear depending on the  charge-to-mass ratio patterns. In the case that ${\bm Z}_1$ and ${\bm Z}_2$ are sufficiently colinear, a single region (and its charge conjugate) exist, as shown  in Fig.\,\ref{fig:CH}, left. In this case, all mass fractions $\sigma_{1,2}$ are allowed provided that the two black holes are sufficiently close to extremality. In particular, the extremal black hole can  split {symmetrically} into two near-extremal black holes of same mass, i.e. with $\sigma_1\approx \sigma_2$.

In contrast, if ${\bm Z}_1$ and ${\bm Z}_2$ are not colinear enough, two disconnected regions (and their charge conjugate) exist, as shown  in Fig.\,\ref{fig:CH}, right. In that case, the kinematic configurations are more restricted. An extremal black hole can only decay into a near-extremal black hole together with a black hole that  is either  very non-extremal or has very small mass fraction.
Hence the decay is necessarily {asymmetric} in this case. 
For example,
an extremal black hole with positive charges can decay into one black hole with  $Z_2\sim \bar Z_2$ together with one satisfying either $Z_1\ll \bar Z_1$ or $\sigma_1\ll1$. 


\section{Magnetic Black Holes Beyond Maxwell} 
\label{app:BH}

\subsection{Solving the Field Equations}

\label{app:FE}

From the effective action \eqref{eq:Gamma_F} we obtain 
the Einstein equation
$    G_{\mu \nu} = \kappa^2 T_{\mu \nu}$
with 
\begin{align}
    T_{\mu \nu} &= -\frac{2}{\sqrt{-g}} \frac{\delta (\sqrt{-g}\mathcal{L}_{F})}{\delta g^{\mu \nu}} = - 4\frac{\partial \mathcal{L}_{F}}{\partial F^2} F_{\mu \lambda}F\indices{_\nu ^\lambda}- 4\frac{\partial \mathcal{L}_{F}}{\partial F \tilde F} F_{\mu \lambda} \tilde F\indices{_\nu ^\lambda} + g_{\mu \nu} \mathcal{L}_{F}\,. 
\end{align}
The field strength equation of motion is
\begin{equation}
    \nabla_\mu \left( \frac{\partial \mathcal{L}_{F}}{\partial F^2} F^{\mu \nu} + \frac{\partial \mathcal{L}_{F}}{\partial (F\Tilde{F})}\Tilde{F}^{\mu \nu}\right) = 0\,.
\end{equation}
Suppose the black hole has magnetic charge $Q$, that generates a radial magnetic field $B_i = (B(r), 0, 0)$. 
Using $B_i = -\frac{1}{2}\epsilon_{i j k}F^{j k}$, we found
\begin{align}
    F^{\theta \phi} = \frac{B(r)}{\sqrt{-g}}\,,\;\;\;\;\;\;\; \Tilde{F}_{t r} = B(r)\,,\;\;\;\;\;\;\;
    F^2 = -\frac{2 r^4 \sin^2{\theta}}{g} (B(r))^2\,,\;\;\;\;\;\;\;
    F \Tilde{F} = 0\,.
\end{align}
Since $T\indices{^t_t} = T\indices{^r _r} = \mathcal{L}_{F}$, we have 
\begin{equation}
    G\indices{^t_t} - G\indices{^r_r} = \frac{f_t(r)}{r}\frac{d}{dr}\left(\frac{f_r(r)}{f_t(r)} \right) = 0\,,
\end{equation}
which implies  $f_t(r) = f_r(r) = f(r)$. The solution for Einstein equation becomes straightforward,
\begin{align}
    G\indices{^t _t} &= \frac{1}{r^2}\left[\frac{d(r f(r))}{dr}- 1\right] =\kappa^2 \mathcal{L}_{F}(2 B^2(r),0)\,,
\end{align}
which gives \eqref{eq:f_gen}. 

\subsection{Properties}

\label{app:Properties}

 Consider the metric \eqref{eq:metric_gen}. 
Using that a static metric has a Killing vector associated with the time symmetry $K^\mu = (1, 0, 0, 0)$, the total mass and magnetic charge are calculated by an integral at spatial infinity,
\begin{align}
    M_{\displaystyle \circ} &= \frac{4 \pi}{\kappa^2}\int_{\partial \Sigma} d^2 x \sqrt{\gamma^{(2)}} n^\mu \sigma^\nu \nabla_\mu K_\nu = \lim_{r\to\infty}\frac{4 \pi r^2 f_t^\prime(r)}{\kappa^2}\sqrt{\frac{f_r(r)}{f_t(r)}} = 4 \pi M \label{eq:mass_parameter}\\
   e Q_{\displaystyle \circ} &= - \int_{\partial \Sigma} d^2 x \sqrt{\gamma^{(2)}} n^\mu \sigma^\nu \tilde F_{\mu \nu} = \lim_{r\to\infty} 4 \pi r^2 \sqrt{\frac{f_r(r)}{f_t(r)}}B_r(r) = 4\pi e Q\,.
\end{align} 
The $n^\mu$ vector  is normal to constant time slices and $n_\mu n^\mu = -1$. The  $\sigma^\mu$ vector is normal to the two-sphere and $\sigma_\mu \sigma^\mu = 1$.

When the gravity sector is pure GR, we have $f_t=f_r\equiv f$ as shown in App.\,\ref{app:FE}. Therefore for any UV completion of the Maxwell sector we have 
\be
 M_{\displaystyle \circ} = \lim_{r\to\infty}\frac{4 \pi r^2 f^\prime(r)}{\kappa^2}\,.
 \ee
Using the solution \eqref{eq:f_gen} we {obtain} 
\begin{equation}
    r^2f'(r) =   \kappa^2 M+ \kappa^2\int_r^\infty dr^\prime r^{\prime 2} \mathcal{L}_F(r) +r^3 {\cal L}_F(r)
    \,.
\end{equation}
Using the assumption that ${\cal L}_F(r)\equiv \mathcal{L}_{F}(2 B^2(r),0)$ is analytical in $B^2$ and vanishes at the origin,  the $r\to \infty$ limit relates the Komar mass $ M_{\displaystyle \circ}$ to the $M$ parameter as $ M_{\displaystyle \circ}=4\pi M$.

\bibliographystyle{JHEP}
\bibliography{biblio}

\end{document}